\documentclass[twocolumn,prc,superscriptaddress,unsortedaddress,showpacs,preprintnumbers,amsmath,amssymb]{revtex4}

\usepackage{graphicx}
\usepackage{dcolumn}

\def\beq{\begin{equation}}
\def\eeq{\end{equation}}
\def\bea{\begin{eqnarray}}
\def\eea{\end{eqnarray}}

\def\fun#1#2{\lower3.6pt\vbox{\baselineskip0pt\lineskip.9pt
  \ialign{$\mathsurround=0pt#1\hfil##\hfil$\crcr#2\crcr\sim\crcr}}}

\begin{document}
\preprint{}

\title{Alpha-decay half-lives and $Q_{\alpha}$ values of superheavy nuclei}

\author{Jianmin Dong}
\email[Electronic address: ]{djm4008@126.com} \affiliation{Institute
of Modern Physics, Chinese Academy of Sciences, Lanzhou 730000,
China} \affiliation{Graduate University of Chinese Academy of
Sciences, Beijing 100049, China}
 \affiliation{School of Nuclear
Science and Technology, Lanzhou University, Lanzhou 730000, China}
\author{Wei Zuo}
\affiliation{Institute of Modern Physics, Chinese Academy of
Sciences, Lanzhou 730000, China}
 \affiliation{School of Nuclear
Science and Technology, Lanzhou University, Lanzhou 730000, China}
\author{Jianzhong Gu}
\affiliation{China Institute of Atomic Energy, P. O. Box 275(18),
Beijing 102413, China}
\author{Yanzhao Wang}
\affiliation{China Institute of Atomic Energy, P. O. Box 275(18),
Beijing 102413, China}
\author{Bangbao Peng}
\affiliation{China Institute of Atomic Energy, P. O. Box 275(18),
Beijing 102413, China}

\date{\today}
\begin{abstract}
The $\alpha$-decay half-lives of recently synthesized superheavy
nuclei (SHN) are investigated based on a unified fission model (UFM)
where a new method to calculate the assault frequency of
$\alpha$-emission is used. The excellent agreement with the
experimental data indicates the UFM is a useful tool to investigate
these $\alpha$-decays. It is found that the half-lives become more
and more insensitive to the $Q_{\alpha}$ values as the atomic number
increases on the whole, which is favorable for us to predict the
half-lives of SHN. In addition, a formula is suggested to compute
the $Q_{\alpha}$ values for the nuclei with $Z\geq 92$ and
$N\geq140$ with a good accuracy, according to which the long-lived
SHN should be neutron rich. With $Q_{\alpha}$ values from this
formula as inputs, we predict the half-lives of isotopes of $Z=117$,
which may be useful for experimental identification in the future.
\end{abstract}

\pacs{27.90.+b, 21.10.Tg, 23.60.+e}

\maketitle
\section{Introduction}\label{intro}\noindent

Syntheses of superheavy nuclei (SHN) becomes an active and exciting
field in modern nuclear physics. Up to now SHN with $Z = 104-118$
except $Z = 117$ have been synthesized in experiment. Superheavy
elements allow nuclear physicists to explore concepts such as magic
numbers and the island of stability, which help us understand the
nuclear structure properties in superheavy region. In recent
experiments on SHN, on the one hand, $\alpha$-decay is indispensable
to the identification of new elements via the observation of
$\alpha$-decay from an unknown parent nucleus to a known daughter
one since the dominant decay mode for SHN is $\alpha$-decay. On the
other hand, experimentalists need the half-life values to design the
experiments. Moreover, measurements on the $\alpha$-decays provide
reliable information on nuclear structure, such as ground state
energies, ground state half-lives, nuclear spins and parities, shell
effects, nuclear deformation and shape coexistence
\cite{RZZ1,HH,PEH,RGL,Audi,TN,DS,APL,CX0}. Therefore, as one of the
most important decay channels for unstable nuclei, $\alpha$-decay
has been extensively investigated both experimentally and
theoretically. From the theoretical point of view, $\alpha$-decay is
regarded as an $\alpha$ particle tunneling through a potential
barrier between an $\alpha$ particle and a daughter nucleus, and
many theoretical models have been applied to investigate the
$\alpha$-decay, such as the cluster model
\cite{CX,XC3,Buck1,Buck2,XXX,D1,D2}, generalized liquid drop model
(GLDM) \cite{Zh1,GRHF,RG1,RG2,RG3,GLDM,DONG}, density-dependent M3Y
(DDM3Y) effective interaction \cite{csp1,csp2,csp3,csp4,csp5} and
coupled channel approach \cite{DD0,DD1}. Some physically plausible
formulas also have been employed to calculate the $\alpha$-decay
half-lives directly \cite{N1,N2,N3}. A unified fission model (UFM)
has been employed to study the proton radioactivity by
Balasubramaniam and Arunachalam \cite{UF1}, and it was used to
extract the preformation factor of cluster in cluster radioactivity
in our previous work \cite{WO}. In this work, the UFM \cite{UF} is
used to study the $\alpha$-decay, in which the assault frequency is
treated with a new approach.

It is well known that the most important decay parameters for
$\alpha$-decay of SHN are the $Q_{\alpha}$ value as well as the
half-life, and $Q_{\alpha}$ value is a key factor for the
$\alpha$-decay half-life calculation. The half-life is extremely
sensitive to the $Q_{\alpha}$ value and an uncertainty of 1 MeV in
$Q_{\alpha}$ corresponds to an uncertainty of $\alpha$-decay
half-life ranging from $10^{3}$ to $10^{5}$ times for the heavy
element region \cite{M1}. Therefore, an accurate formula of
$Q_{\alpha}$ value is crucial for the half-life prediction. However,
the calculated $Q_{\alpha}$ value with the extant methods is
difficult to achieve a good accuracy. Therefore, we derive an
expression of $Q_{\alpha}$ value based on the liquid drop model,
which can be used as an input to quantitatively predict the
half-lives of unknown nuclei.

\section{Theoretical framework of the UFM}\label{intro}\noindent
The half-life of a parent nucleus decaying via $\alpha$ emission can
be calculated by means of the WKB barrier penetration probability.
In the UFM, the decay constant is simply defined as $\lambda =\nu
_{0}P$ and half-life can be obtained by $T=\ln 2/\lambda $. Here
$\nu _{0}$ is the assault frequency which will be addressed in
detail later. The barrier penetrability $P$ is given by
\begin{equation}
P=\exp \left[ -\frac{2}{\hbar
}\int_{R_{\text{in}}}^{R_{\text{out}}}\sqrt{2\mu \left(
V(r)-Q_{\alpha}\right) }dr\right],
\end{equation}
where $R_{\text{in}}$ and $R_{\text{out}}$ are incoming and outgoing
points with $V(R_{\text{in}})=V(R_{\text{out}})=Q_{\alpha}$. The
potential $V(r)$ is composed of the repulsive long range Coulomb
potential, the attractive short range nuclear proximity potential
and the centrifugal potential for $r\geq R_{1}+R_{2}$, but for $r<
R_{1}+R_{2}$, $V(r)$ is parameterized simply as a polynomial. Here
$R_{0}$, $R_{1}$ and $R_{2}$ are the radii of the parent nucleus,
daughter one and emitted particle respectively, which are given by
\cite{UF,PPP}
\begin{equation}
R_{i}=(1.28A_{i}^{1/3}-0.76+0.8A_{i}^{-1/3})~\text{fm, \ \ }\
i=0,1,2.
\end{equation}
In a word, the potential $V(r)$ takes the form
\begin{equation}V(r)=
\begin{cases}
{a_{0}+a_{1}r+a_{2}r^{2}} & \text{for $R_{0}\leq r<R_{1}+R_{2}$}\\
 V_{p}(r)+V_{l}(r)+ \frac{Z_{1}Z_{2}e^{2}}{r}& \text{for $r\geq
R_{1}+R_{2}$},\\
\end {cases}
\end{equation}
where $Z_{1}$ and $Z_{2}$ are the charge numbers of the emitted
particle and daughter nucleus, respectively. The coefficients
$a_{0}$, $a_{1}$, $a_{2}$ in the polynomial can be determined by the
following boundary conditions

(1) At $r=R_{0}$, $V(r)=Q_{\alpha}$;

(2) At $r=R_{1}+R_{2}$, $V(r)=V(R_{1}+R_{2})$;

(3) At $r=R_{1}+R_{2}$,
$\frac{dV(r)}{dr}=\frac{dV(r)}{dr}|_{r=R_{1}+R_{2}}$.\\
The third condition ensures the smooth potential curve, which is
different from the previous UFM. $R_{\text{in}}$ is the internal
turning point with $V(r)=Q_{\alpha}$, which differs from Refs.
\cite{HF1,WO} where the barrier penetration probabilities were
calculated from two touching spheres ($R_{\text{in}}=R_{1}+R_{2}$).
Therefore, the formation process of cluster or $\alpha$-particle was
not be taken into account and hence one can extract the preformation
factor by combining with the experimental half-life. In our
calculations with the present potential barrier, however, the
penetrability has been evaluated from $R_{\text{in}}$, hence the
process of formation of $\alpha$-particle has been considered to a
great extent, as has been pointed out in Ref. \cite{KKK}: The
preformation probability can be calculated within a fission model as
a penetrability of the internal part of the barrier, which
corresponds to still overlapping fragments. $R_{\text{out}}$ is
given by
\begin{equation}
R_{\text{out}}=\frac{Z_{1}Z_{2}e^{2}}{2Q_{\alpha}}+\sqrt{\left( \frac{Z_{1}Z_{2}e^{2}%
}{2Q_{\alpha}}\right) ^{2}+\frac{l(l+1)\hbar ^{2}}{2\mu
Q_{\alpha}}}.
\end{equation}
$V_{p}(r)$ in the potential is the nuclear proximity potential
taking the form
\begin{equation}
V_{p}(r)=4\pi \frac{C_{1}C_{2}}{C_{1}+C_{2}}\gamma b\Phi (s),
\label{P}
\end{equation}
where S\"{u}smann central radii are $C_{i}=R_{i}-b^{2}/R_{i}$ and
$b=0.99$ fm is the surface width. The nuclear surface tension
coefficient $\gamma$ is given as
\begin{equation}
\gamma =0.9517\left[ 1-1.7826\left( \frac{N-Z}{A}\right) ^{2}\right]
\text{ \ \ MeV}\cdot \text{fm}^{-2},
\end{equation}
where $N$, $Z$ and $A$ represent the neutron, proton and mass
numbers of the parent nucleus. The universal function $\Phi(s)$ is
determined by the following formula \cite{UF,PPP}
\begin{equation}\Phi (s)=
\begin{cases}
{-\frac{1}{2}(s-2.54)^{2}-0.0852(s-2.54)^{3}}, & s\leq 1.2511\\
 -3.437\exp \left( -\frac{s}{0.75}\right),& s>1.2511\\
\end {cases}
\end{equation}
where $s=(r-C_{1}-C_{2})/b$ is the overlap distance in units of $b$
between the colliding surfaces.

We propose a new approach to deal with the assault frequency
phenomenologically. Assuming that the $\alpha$ particle vibrates in
a harmonic oscillator potential $V(r)=-V_{0}+\frac{1}{2}\mu \omega
^{2}r^{2}$ with a classical frequency $\omega$ and a reduced mass
$\mu$ after formation, by employing the Virial theorem, we obtain
\begin{equation}
\mu \omega ^{2}\overline{r^{2}}=(2n_{r}+l+\frac{3}{2})\hbar \omega,
\end{equation}
where $n_{r}$ and $l$ are the radial quantum number (corresponding
to the number of nodes) and angular momentum quantum number,
respectively. $\sqrt{\overline{r^{2}}}=<\psi |r^{2}|\psi >^{1/2}$ is
the root-mean-square radius of $\alpha$ particle distribution in
quantum mechanics and that it equals to the rms radius
$R_{\text{n}}$ of nucleus is assumed here. It is farfetched that the
assault frequency is understood with a classical method that the
$\alpha$ particle moving back and forth inside the nucleus due to
the wave properties of the $\alpha$ particle. We identify the
oscillation frequency $\nu_{0}$ with the assault frequency, which is
related to the oscillation frequency $\omega$
\begin{equation}
\nu _{0}=\frac{\omega }{2\pi }=\frac{(2n_{r}+l+\frac{3}{2})\hbar
}{2\pi \mu R_{\text{n}}^{2}}=\frac{(G+\frac{3}{2})\hbar }{1.2\pi \mu
R_{0}^{2}}. \label{A}
\end{equation}
The relationship of $R_{n}^{2}=\frac{3}{5}R_{0}^{2}$ \cite{R} is
used here. $G=2n_{r}+l$ is the principal quantum number. For
$\alpha$-decay, we take the form as in Ref. \cite{CX}
\begin{equation}G=2n_{r}+l=
\begin{cases}
22, & N>126\\
20,& 82<N\leq126\\
18,&N\leq82.
\end {cases}
\end{equation}
The order of magnitude of $\nu _{0}$ is $10^{21}$ s$^{-1}$ for
$\alpha$-decay. As have been pointed out in Ref. \cite{JC}, the
quantum number $G$ can have an uncertainty of 2 due to the simple
application of Wildermuth rule to heavy nuclei that involve shell
mixtures, but not serious.

The calculations are performed in the framework of spherical shape,
which is partly equivalent to averaging the deformed potential to a
spherical case. Recently, some authors investigated the
$\alpha$-decay in the framework of the deformed version of the
$\alpha$-decay model. We would point out that the centrifugal
barrier should not take the form of $\hbar ^{2}(l+1/2)^{2}/(2\mu
r^{2})$ because $l$ is not a good quantum number for the deformed
potential.

\section{Half-lives of the newly synthesized SHN}\label{intro}\noindent

The $\alpha$-decay half-lives of SHN calculated with the UFM using
the experimental $Q_{\alpha}$ values and without considering the
centrifugal barrier are given in Table I. The results obtained with
the DDM3Y effective interaction and the GLDM also have been shown
for comparison. The results from the UFM are in fair agreement with
the experimental data indicating that the UFM taking account of the
assault frequency with the phenomenological method is a useful tool
to investigate the half-lives of $\alpha$-decay when the
experimental $Q_{\alpha}$ values are given. The DDM3Y effective
interaction overestimates but GLDM underestimates the half-lives on
the whole. There is no doubt that the DDM3Y interaction and GLDM are
very successful due to the appropriate considerations in the
microscopic level in the DDM3Y interaction and the quasi-molecular
shape in the GLDM. The deviations might result from the fact that
empirical assault frequencies they used are too rough. The UFM is
quite simple compared to the GLDM and DDM3Y interaction, but
provides the excellent results. Another obvious advantage is that
the proximity potential for proton, $\alpha$ and cluster emission
can be written in a unified manner, which means these different
decay modes can be easily treated in a unified framework. For some
nuclei belonging to $^{282}$113, $^{280}$111 and $^{279}$111
$\alpha$-decay chains, the half-lives from the UFM as well as other
models are underestimated by a few times possibly due to the nonzero
angular momentum transfers, which reduce the tunneling probability
and increase the half-life. However, as no experimental evidence is
available for the spin-parity of the levels involved in the decay,
we have not included the centrifugal barrier in the calculations.

Recently, the new isotope $^{263}$Hs has been produced in the
reaction $^{208}$Pb($^{56}$Fe, n)$^{263}$Hs at the 88-Inch Cyclotron
of the Lawrence Berkeley National Laboratory \cite{Hs}. There are
three $\alpha$-particle energy groups at $E_{\alpha}=10.57\pm0.06$,
$10.72\pm0.06$, and $10.89\pm0.06$ MeV observed in experiment. The
calculated half-life of $0.24^{+0.10}_{-0.07}$ ms assuming
$E_{\alpha}=10.57\pm0.06$ MeV ($Q_{\alpha}=10.78\pm0.06$ MeV) is
closest to the experimental data of $0.74^{+0.48}_{-0.21}$ ms which
indicates the group of the $E_{\alpha}=10.57\pm0.06$ MeV is perhaps
the dominant transition among the three groups.

It is an interesting phenomenon that most of odd-A or odd-odd SHN
are longer-lived than the even-even ones around them which perhaps
indicates the stability of odd-A or odd-odd nuclei over the
even-even ones. On the one hand, the small preformation probability
could prolong the $\alpha$-decay half-life since the dominant decay
mode for SHN is $\alpha$-decay. On the other hand, the possible
centrifugal barrier reduces tunneling probability and hence
increases the lifetime. This problem needs to be studied further.
The odd-A isotopes of all the elements with $Z=116,114,112,110$ and
$108$, which lie in the neighborhood of the even-even isotopes, have
been observed. This may suggest that $^{293}$118 and $^{295}$118 can
be the good candidates to be synthesized in laboratory since the new
element $^{294}$118 has been synthesized.

\begin{figure}[htbp]
\begin{center}
\caption{$K$ value as a function of atomic number.}
\end{center}
\end{figure}

\begin{figure}[htbp]
\begin{center}
\caption{$K$ values of Po, Rn, Ra and Th isotopes as a function of
neutron number.}
\end{center}
\end{figure}

It is well known that the $Q_{\alpha}$ value is a crucial quantity
to determine the $\alpha$-decay half-life. Up to now, however, there
has been nearly no approach that can provide an accurate
$Q_{\alpha}$ value theoretically with deviation less than 0.5 MeV,
leading to the prediction of half-life with a good accuracy a very
difficult work. Here we introduce a quantity
\begin{equation}
K=\left\vert \frac{\partial \left[ \log _{10}T_{\alpha }(\text{s})\right] }{%
\partial Q_{\alpha}}\right\vert,
\end{equation}
which describes the $Q_{\alpha}$ value dependence of $\alpha$-decay
half-life. To show the behavior of $K$ values more obviously, we
calculate the $K$ values including heavy nuclei ranging from $Z=62$
to $Z=118$, and show the results in Fig. 1. One could notice that,
the $K$ value decreases with increasing of the atomic number $Z$ on
the whole. This indicates the half-life becomes more and more
insensitive to $Q_{\alpha}$ value. For instance, the increase of
$Q_{\alpha}$ value by 0.4 MeV leads to the half-life decrease by
only one order of magnitude for $^{294}$118, but five orders of
magnitude for $^{147}_{62}$Sm. This is an advantage for us to
predict the $\alpha$-decay half-lives of SHN since they are not so
sensitive to $Q_{\alpha}$ value as for medium-heavy nuclei. For some
nuclei near the $Z=82$ closure shell, the $K$ values are low because
they are strongly affected by the shell effect. We present the $K$
values of even-even Po, Rn, Ra and Th isotopes in Fig. 2. One can
find that the larger the atomic number of an element, the lower the
$K$ values, which further confirms what we have discussed above. The
$K$ value changes smoothly before $N=126$, but decreases sharply
from $N=126$ to $N=128$, and increases rapidly after $N=128$ with
increasing of neutron number, indicating the shell effect plays an
important role in the behavior of $K$ value. This fact suggests that
for a given superheavy element, the isotopes at the beginning of the
closed shell are more insensitive to $Q_{\alpha}$ values.

\begin{table*}[h]
\label{table1} \caption{Comparisons between the experimental and
theoretical $\alpha$-decay half-lives of recently synthesized
superheavy nuclei. The experimental data are from Ref. \cite{YY1}
and the latest data are listed.}
\begin{ruledtabular}
\begin{tabular}{llllllll}
Nucleus& $Q^{\text{expt}}_{\alpha}$(MeV)&  $T^{\text{expt}}_{\alpha}$&  $T^{\text{UFM}}_{\alpha}$& $ T^{\text{DDM3Y}}_{\alpha}$ \cite{csp1}& $T^{\text{GLDM}}_{\alpha}$ \cite{GRHF,DONG}  \\
\hline
  $^{294}$118&$11.81\pm0.06$& $0.89^{+1.07}_{-0.31}$ ms &$0.59^{+0.23}_{-0.16}$ ms &$0.66^{+0.23}_{-0.18}$ ms &$0.15^{+0.05}_{-0.04}$ ms   \\
  $^{293}$116&$10.67\pm0.06$& $53^{+62}_{-19}$ ms &$93.2^{+42.2}_{-28.8}$ ms  &$206^{+90}_{-61}$ ms &$22.81^{+10.22}_{-7.06}$ ms       \\
  $^{292}$116&$10.80\pm0.07$& $18^{+16}_{-6}$ ms  &$43.5^{+23.2}_{-15.0}$ ms  &$39^{+20}_{-13}$ ms  &$10.45^{+5.65}_{-3.45}$ ms        \\
  $^{291}$116&$10.89\pm0.07$&$18^{+22}_{-6}$ ms &$26.2^{+13.8}_{-8.9}$ ms &$60.4^{+30.2}_{-20.1}$ ms &$6.35^{+3.15}_{-2.08}$ ms     \\
 $^{290}$116&$11.00\pm0.08$& $7.1^{+3.2}_{-1.7}$ ms &$14.1^{+8.6}_{-5.3}$ ms   &$13.4^{+7.7}_{-5.2}$ ms &$3.47^{+1.99}_{-1.26}$ ms  \\
  $^{288}$115  &$10.61\pm0.06$&87 $^{+105}_{-30}$ ms & $72.2^{+32.7}_{-22.4}$ ms    & $410.5^{+179.4}_{-122.7}$ ms  &94.7$^{+41.9}_{-28.9}$ ms   \\
  $^{287}$115  & $10.74\pm0.09$ &   32$^{+155}_{-14}$ ms & $33.7^{+24.8}_{-14.2}$ ms   &   $51.7^{+35.8}_{-22.2}$  ms   &46.0$^{+33.1}_{-19.1}$ ms \\
  $^{289}$114& $9.96\pm0.06$& $2.7^{+1.4}_{-0.7}$ s &$2.05^{+1.03}_{-0.68}$ s &$3.8^{+1.8}_{-1.2}$ s  &$0.52^{+0.25}_{-0.17}$ s    \\
  $^{288}$114&$10.09\pm0.07$& $0.8^{+0.32}_{-0.18}$ s & $0.89^{+0.53}_{-0.33}$ s &$0.67^{+0.37}_{-0.27}$ s &$0.22^{+0.12}_{-0.08}$ s   \\
  $^{287}$114&$10.16\pm0.06$& $0.48^{+0.16}_{-0.09}$ s& $0.58^{+0.28}_{-0.19}$ s  &$1.13^{+0.52}_{-0.40}$ s  &$0.16^{+0.08}_{-0.05}$ s  \\
  $^{286}$114&$10.33\pm0.06$& $0.13^{+0.04}_{-0.02}$ s & $0.20^{+0.09}_{-0.06}$ s&$0.16^{+0.07}_{-0.05}$ s &$0.05^{+0.02}_{-0.02}$ s   \\
  $^{284}$113  &$10.15\pm0.06$& 0.48$^{+0.58}_{-0.17}$ s &$0.30^{+0.14}_{-0.10}$ s &1.55$^{+0.72}_{-0.48}$ s&0.43$^{+0.21}_{-0.13}$ s       \\
  $^{283}$113  & $10.26\pm0.09$&100$^{+490}_{-45}$ ms  & $153.2^{+120.6}_{-66.8}$ ms   &201.6$^{+164.9}_{-84.7}$ms&222$^{+172}_{-96}$ ms     \\
$^{282}$113&$10.83\pm0.08$\tablenotemark[1]& $73^{+134}_{-29}$ ms &$4.8^{+2.9}_{-1.8}$ ms &--&$7.8^{+4.6}_{-2.8}$ ms  \\
   $^{285}$112&$9.29\pm0.06$& $34^{+17}_{-9}$ s&    $48.0^{+26.9}_{-17.1}$ s  &$75^{+41}_{-26}$ s &$13.22^{+7.25}_{-4.64}$ s  \\
  $^{283}$112& $9.67\pm0.06$& $3.8^{+1.2}_{-0.7}$ s &$3.4^{+1.8}_{-1.1}$ s  &$5.9^{+2.9}_{-2.0}$ s &$0.95^{+0.48}_{-0.32}$ s    \\
  $^{280}$111  & $9.87\pm0.06$&  3.6 $^{+4.3}_{-1.3}$  s  &$0.41^{+0.20}_{-0.14}$ s &1.9$^{+0.9}_{-0.6}$ s  &0.69$^{+0.33}_{-0.23}$ s     \\
  $^{279}$111  & $10.52\pm0.16$&170$^{+810}_{-80}$ ms &$6.8^{+11.3}_{-4.2}$ ms   &9.6$^{+14.8}_{-5.7}$ ms&12.4$^{+19.9}_{-7.6}$ ms      \\
$^{278}$111&$10.89\pm0.08$\tablenotemark[1]& $4.2^{+7.5}_{-1.7}$ ms&$0.79^{+0.47}_{-0.29}$ ms  &-- &$1.5^{+0.9}_{-0.5}$ ms \\
  $^{279}$110&$9.84\pm0.06$& $0.20^{+0.05}_{-0.04}$ s& $0.22^{+0.11}_{-0.07}$ s  &$0.40^{+0.18}_{-0.13}$ s &$0.08^{+0.04}_{-0.02}$ s  \\
  $^{276}$109  & $9.85\pm0.06$& 0.72$^{+0.87}_{-0.25}$ s & $0.10^{+0.05}_{-0.03}$ s&0.45$^{+0.23}_{-0.14}$ s&0.19$^{+0.08}_{-0.06}$ s      \\
 $^{275}$109  & $10.48\pm0.09$& 9.7$^{+46}_{-4.4}$  ms &$1.97^{+1.42}_{-0.82}$ ms &2.75$^{+1.85}_{-1.09}$ ms&4.0$^{+2.8}_{-1.6}$ ms       \\
$^{274}$109&$9.95\pm0.10$\tablenotemark[1]& $440^{+810}_{-170}$ ms &$55.6^{+51.3}_{-26.4}$ ms &--&$108^{+96}_{-51}$ ms   \\
  $^{275}$108&$9.44\pm0.06$& $0.19^{+0.22}_{-0.07}$ s &$0.70^{+0.36}_{-0.24}$ s &$1.09^{+0.61}_{-0.35}$ s &$0.27^{+0.16}_{-0.10}$ s    \\
  $^{272}$107  & $9.15\pm0.06$&  9.8$^{+11.7}_{-3.5}$  s & $2.53^{+1.38}_{-0.89}$ s  &10.1$^{+5.4}_{-3.4}$ s&5.12$^{+3.19}_{-1.58}$  s     \\
$^{270}$107&$9.11\pm0.08$\tablenotemark[1]& $61^{+292}_{-28}$ s &$3.6^{+2.9}_{-1.6}$ s &--&$7.7^{+6.1}_{-3.3}$ s  \\
  $^{271}$106&$8.67\pm0.08$& $1.9^{+2.4}_{-0.6}$ min &$0.64^{+0.56}_{-0.30}$ min  &$0.86^{+0.71}_{-0.39}$ min &$0.33^{+0.28}_{-0.16}$ min   \\
\end{tabular}
\footnotetext[1] {$Q_{\alpha}$ values are calculated using the
measured $\alpha$ kinetic energies. The electron shielding
corrections have been taken into account.}
\end{ruledtabular}
\end{table*}

\section{Formula of $Q_{\alpha}$ value for nucleus with $Z\geq
92$ and $N\geq140$}\label{intro}\noindent

Let us turn to the $Q_{\alpha}$ value of SHN. The starting point is
the local formula of binding energy for the nuclei with $Z\geq 90$
and $N\geq140$ \cite{DTK}:
\begin{eqnarray}
B(Z,A) &=&a_{v}A-a_{s}A^{2/3}-a_{c}Z^{2}A^{-1/3}-a_{a}\left( \frac{A}{2}%
-Z\right) ^{2}A^{-1}\nonumber \\
&&+a_{p}A^{-1/2}+a_{6}\left\vert A-252\right\vert /A-a_{7}\left\vert
N-152\right\vert /N\nonumber \\
&&+a_{8}\left\vert N-Z-50\right\vert /A.\label{AA}
\end{eqnarray}
This formula can achieve a high accuracy for binding energy.
However, when it is employed to calculate the $Q_{\alpha}$ value in
terms of mass deficit for SHN, the large deviation can be found, as
shown in Table III in Ref. \cite{DTK}. It might be feasible to
deduce a more accurate formula for $Q_{\alpha}$ with Eq. (\ref{AA})
because some terms for parent and daughter nuclei may cancel out
approximately and contains few parameters. Here we only focus on the
nuclei with $Z\geq 92$ and $N\geq140$. According to Eq. (\ref{AA}),
the $Q_{\alpha}$ value can be written as
\begin{eqnarray}
Q_{\alpha } &=&B(\alpha )+B(Z-2,A-4)-B(Z,A)  \notag \\
&=&B(\alpha )+\left[ a_{v}(A-4)-a_{v}A\right] +\left[
a_{s}A^{2/3}-a_{s}(A-4)^{2/3}\right]   \notag \\
&&+\left[ -a_{c}(Z-2)^{2}(A-4)^{-1/3}+a_{c}Z^{2}A^{-1/3}\right] +  \notag \\
&&\left[ -a_{a}\left( \frac{N-Z}{2}\right) ^{2}(A-4)^{-1}+a_{a}\left( \frac{%
N-Z}{2}\right) ^{2}A^{-1}\right]   \notag \\
&&+\left[ a_{p}\delta \left( A-4\right) ^{-1/2}-a_{p}\delta
A^{-1/2}\right]
\notag \\
&&+\left[ a_{6}\frac{\left\vert A-256\right\vert }{A-4}-a_{6}\frac{%
\left\vert A-252\right\vert }{A}\right]   \notag \\
&&+\left[ -a_{7}\frac{\left\vert N-154\right\vert }{N-2}+a_{7}\frac{%
\left\vert N-152\right\vert }{N}\right]   \notag \\
&&+\left[ a_{8}\frac{\left\vert N-Z-50\right\vert }{A-4}-a_{8}\frac{%
\left\vert N-Z-50\right\vert }{A}\right]   \notag \\
&\approx &B(\alpha )-4a_{v}+\frac{8}{3}a_{s}A^{-1/3}+\frac{4}{3}%
a_{c}ZA^{-4/3}(3A-Z)  \notag \\
&&-a_{a}\left( \frac{N-Z}{A}\right) ^{2}+2a_{p}A^{-3/2}+  \notag \\
&&a_{6}\left[ \frac{\left\vert A-256\right\vert
}{A-4}-\frac{\left\vert
A-252\right\vert }{A}\right]   \notag\\
&&a_{7}\left[ \frac{\left\vert N-152\right\vert
}{N}-\frac{\left\vert
N-154\right\vert }{N-2}\right] +4a_{8}\frac{\left\vert N-Z-50\right\vert }{%
A(A-4)}.
\end{eqnarray}
In the process of deduction, the Taylor Expansion was used. As a
constant, $B(\alpha )$ is the binding energy of $\alpha$-particle.
According to the parameters provided by Ref. \cite{DTK}, we can
estimate the contribution from each term. It is found that the
paring energy ($a_{p}$ term), the $a_{6}$ and the $a_{8}$ terms
contribute very little to the $Q_{\alpha}$ value and can be
neglected. The volume energy ($a_{v}$ term) is only a constant and
the surface energy ($a_{s}$ term) can be regarded as a constant
since it varies very little in this local region. The term
$a_{7}\left[ \left\vert N-152\right\vert /N-\left\vert
N-154\right\vert /(N-2)\right] $ simulates the deformed shell effect
of $N=152$. Similarly, we introduce a new term $a_{9}\left[
\left\vert Z-Z_{0}\right\vert /Z-\left\vert Z-Z_{0}-2\right\vert
/(Z-2)\right] $ to simulate the proton shell effect. We find
$Z_{0}=110$ in our fitting procedure latter, which indicates that a
possible shell gap exists at $Z=110$, and we set $Z_{0}=110$ here
beforehand for convenience. Therefore, the above formula can be
simplified further to
\begin{eqnarray}
Q_{\alpha }(\text{MeV}) &=&aZA^{-4/3}(3A-Z)+b\left( \frac{N-Z}{A}\right) ^{2} \nonumber\\
&&+c\left[ \frac{\left\vert N-152\right\vert }{N}-\frac{\left\vert
N-154\right\vert }{N-2}\right]\nonumber  \\
&&+d\left[ \frac{\left\vert Z-110\right\vert }{Z}-\frac{\left\vert
Z-112\right\vert }{Z-2}\right] +e.\label{BB}
\end{eqnarray}

\begin{figure}[htbp]
\begin{center}
\caption{The deviations between the formula (\ref{BB}) and
experimental $Q_{\alpha}$ values for 154 nuclei with $Z\geq 92$ and
$N\geq140$ as a function of proton number.}
\end{center}
\end{figure}

The coefficients above are obtained by fitting the 154 experimental
data with $Z\geq 92$ and $N\geq140$. Some experimental data are
taken from Ref. \cite{AUDI0} and Table I, $^{260}$Bh from \cite{Bh},
$^{237}$Cm from \cite{Cm}, $^{258}$Rf from \cite{Rf}. That $\alpha$
transitions occur from ground states to ground states is assumed for
all decays here. The best fit parameters are
\begin{equation}
\begin{cases}
a=0.9373\text{\ MeV}, \\
b=-99.3027\text{\ MeV}, \\
c=16.0363\text{\ MeV} ,\\
d=-21.5983\text{\ MeV}, \\
e=-27.4530\text{\ MeV}.
\end{cases}%
\end{equation}
The standard and average deviations of the $Q_{\alpha}$ value for
the 154 nuclei are as follows
\begin{eqnarray}
\sqrt{\overline{\sigma ^{2}}} &=&\sqrt{\overset{154}{\underset{i=1}{\sum }}%
\frac{1}{154}\left( Q_{\text{expt}.}^{i}-Q_{\text{cal}.}^{i}\right)
^{2}}=0.183,
\\
\overline{\sigma } &=&\overset{154}{\underset{i=1}{\sum }}\frac{1}{154}%
\left\vert Q_{\text{expt}.}^{i}-Q_{\text{cal}.}^{i}\right\vert
=0.137.
\end{eqnarray}
The little deviation of $Q_{\alpha}$ value confirms Eq. (\ref{BB})
will be very useful for experiments and it only contains five
parameters while Eq. (\ref{AA}) contains eight ones. We plot the
deviations between the Eq. (\ref{BB}) and the experimental
$Q_{\alpha}$ values in Fig. 3. As Ref. \cite{DTK} pointed out,
$N=162$ is a magic number since the systematic deviations between
theoretical and experimental $Q_{\alpha}$ values near $N=164$. In a
completely analogous manner, systematic deviations in Fig. 3 imply
that a possible shell gap exists at $Z=108$ which has been discussed
in many works \cite{shell1,shell2}. From Eq. (\ref{BB}), one can see
that the contributions of the coulomb energy and symmetry energy are
just opposite, the symmetry energy contributing negatively and much
larger than those of shell effects. For long-lived SHN, the
$Q_{\alpha}$ value should be smaller, which means the relatively
larger absolute symmetry energy for a given element. In other words,
long-lived SHN should be neutron rich. The neutron rich SHN is
difficult to produce with the existing facilities. However, with the
upcoming RIB facilities and improved detection techniques, we
believe that such long-lived SHN would be synthesized in the near
future.

For the nuclei with $Z\geq112$, Eq. (\ref{BB}) can give a very good
description, hence Eq. (\ref{BB}) can be used to predict the
$Q_{\alpha}$ value with a good accuracy especially for $Z\geq112$.
The $^{293}$118 and $^{295}$118 may be synthesized in the near
future, the half-lives of which are predicted to be 0.49 ms and 1.99
ms by employing the UFM with Eq. (\ref{BB}) as inputs. The
superheavy element with $Z=117$ has not been observed in experiment
up to now, and some theoretical investigations have been carried out
on it \cite{LZH}. We predict the half-lives of isotopes of $Z=117$
with the $Q_{\alpha}$ value from Eq. (\ref{BB}), and the results are
listed in table II, which may be useful for future experiments.

\begin{table}
\label{table2} \caption{Predicted $\alpha$-decay half-lives of
$Z=117$ isotopes using the UFM with the $Q_{\alpha}$ values from Eq.
(\ref{BB}).}
\begin{ruledtabular}
\begin{tabular}{lllllllllllll}
nuclei & $Q_{\alpha}$(MeV) & $T_{\alpha}^{\text{UFM}}$ &nuclei & $Q_{\alpha}$(MeV) & $T_{\alpha}^{\text{UFM}}$  \\
\hline
$^{288}117$  & 11.94 & 0.17 ms   &  $^{289}117$  & 11.81 & 0.33 ms\\
$^{290}117$  & 11.67 & 0.68 ms   &  $^{291}117$  & 11.54 & 1.34 ms\\
$^{292}117$  & 11.40 & 2.83 ms & $^{293}117$ & 11.27  &  5.73 ms\\
$^{294}117$ & 11.13 &12.5 ms & $^{295}117$ & 10.99 &27.5 ms \\
$^{296}117$ & 10.85 &61.7 ms & $^{297}117$ & 10.71 &0.14 s \\
$^{298}117$ & 10.57 &0.33 s & $^{299}117$ & 10.43 &0.77 s \\
\end{tabular}
\end{ruledtabular}
\end{table}

\section{Summary}\label{intro}\noindent

In summary, the half-lives of $\alpha$-decay for SHN have been
investigated in the framework of a UFM with a new method for assault
frequency. No adjustable parameter has been involved in the
calculations. The results of the present calculations using the UFM
are in excellent agreement with the experimental data. For some
nuclei in $^{282}$113, $^{280}$111 and $^{279}$111 $\alpha$-decay
chains, the half-lives from the UFM together with other models are
underestimated by a few times possibly due to the nonzero angular
momentum transfers. We also find that $Q_{\alpha}$ value dependence
of $\alpha$-decay half-life becomes increasing weaker as the atomic
number increases on the whole, which implies that the uncertainty of
the $\alpha$-decay half-life due to the uncertainty of $Q_{\alpha}$
value is smaller for heavier nuclei and thus it is exactly what we
expect to predict $\alpha$-decay half-life of SHN. And the isotopes
at the beginning of the closed shell are more insensitive to
$Q_{\alpha}$ values. Finally, a local formula was proposed to
calculate the $Q_{\alpha}$ values for the nuclei with $Z\geq 92$ and
$N\geq140$. According to this formula in combination with the
experimental data, the possible proton shell gaps exist at $Z=108$
and 110, and long-lived SHN should be neutron rich. The half-lives
of isotope of $Z=117$ which perhaps will be observed in the near
future, are predicted by using the UFM combing with this formula.

This work is supported by the National Natural Science Foundation of
China (10875151,10575119,10675170,10975190), the Major State Basic
Research Developing Program of China under No. 2007CB815003 and
2007CB815004, the Knowledge Innovation Project(KJCX3-SYW-N2) of
Chinese Academy of Sciences, CAS/SAFEA International Partnership
Program for Creative Research Teams (CXTD-J2005-1).

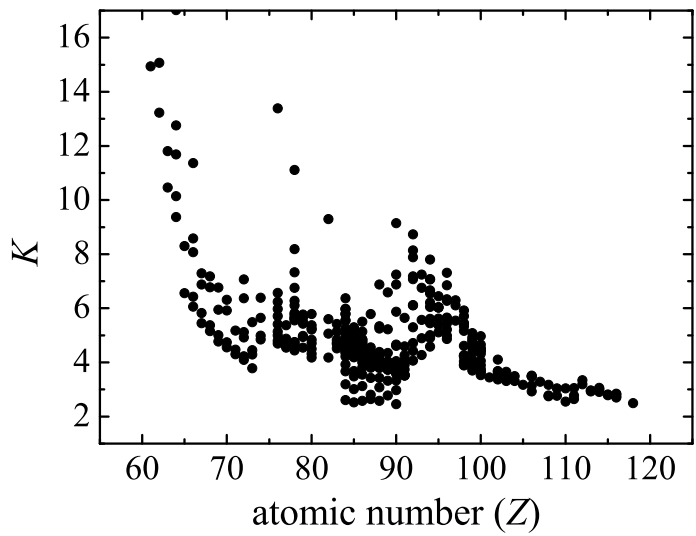
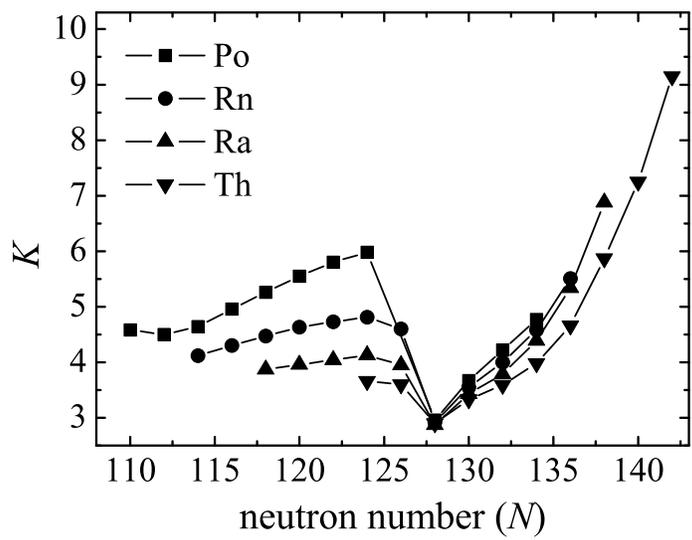
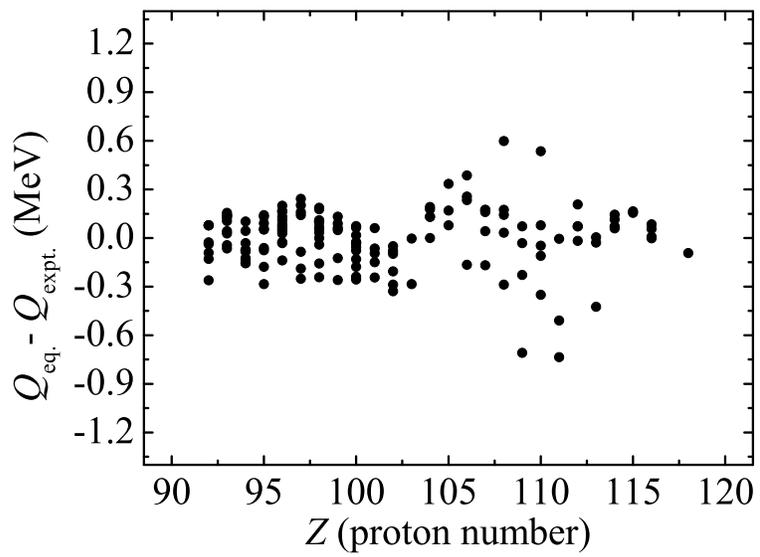
\end{document}